\documentclass[aps,prl,reprint,twocolumn,showpacs]{revtex4}
\usepackage{amsmath}
\usepackage{graphicx}
\usepackage{epstopdf}

\usepackage{float}
\usepackage[caption = false]{subfig}

\begin{document}
\title{Sub Decoherence Time Generation and Detection of Orbital Entanglement}
\author{F. Brange}
\author{O. Malkoc}
\author{P. Samuelsson}
\affiliation{Department of Physics, Lund University, Box 118, SE-221 00 Lund, Sweden}


\begin{abstract}
Recent experiments have demonstrated sub decoherence time control of individual single-electron orbital qubits. Here we propose a quantum dot based scheme for generation and detection of pairs of orbitally entangled electrons on a timescale much shorter than the decoherence time. The electrons are entangled, via two-particle interference, and transferred to the detectors during a single cotunneling event, making the scheme insensitive to charge noise. For sufficiently long detector dot lifetimes, cross-correlation detection of the dot charges can be performed with real-time counting techniques, opening up for an unambiguous short-time Bell inequality test of orbital entanglement.
\end{abstract}

\maketitle

The concept of quantum entanglement has ever since its inception attracted much attention. Initially questioned because of its nonlocal properties, violating local realism \cite{PhysRev.47.777, Bell1964}, entanglement has over the past decades emerged as an indispensable resource for quantum information processing \cite{Nielsen&ChuangI}. Spurred by proposals for electronic spin-based quantum computing \cite{PhysRevA.57.120, PhysRevLett.89.147902}, spin qubit experiments \cite{Science.309.5744, 10.1038/nature05065} and demonstrations of long spin decoherence times \cite{Fischer05062009}, large efforts have been devoted to investigations of spin entanglement in nanostructures. Recent experimental progress comprises entanglement of single-electron \cite{PhysRevLett.107.146801} and two-electron \cite{Science.336.202} spin qubits and splitting \cite{Nature.461.960, PhysRevLett.104.026801, Nat.Comm.3.1165, PhysRevLett.109.157002} of spin-singlet Cooper pairs in hybrid superconducting systems.

In contrast to spin, entanglement between electronic orbital degrees of freedom \cite{PhysRevLett.84.5912, PhysRevLett.91.157002}, such as charge states in quantum dots \cite{PhysicaE.43.730} or edge channels in quantum Hall systems \cite{PhysRevLett.92.026805, PhysRevLett.91.147901, Nature.448.333}, has received limited attention. In particular, orbital entanglement has not been demonstrated experimentally. The key reason is arguably that superpositions of orbital states are sensitive to charge noise, resulting in short decoherence times, of the order of nanoseconds \cite{PhysRevLett.91.226804, PhysRevLett.100.126802, PhysRevLett.105.246804, AppPhysLett.105.063105}. This has led to the widespread view that, despite all-electrical quantum state control and read-out, electronic orbital degrees of freedom cannot be harnessed for quantum information processing. Very recently this view was contested by demonstrations of fast, coherent operations of single-electron orbital qubits on the picosecond timescale \cite{PhysRevB.84.161302, Nature.7.247, Nature.Comm.4.1401}, several orders of magnitude shorter than the decoherence time. These experiments motivate renewed efforts on orbital-based quantum information processing and call for novel schemes to generate and detect orbital entanglement on timescales well below the decoherence time.

\begin{figure}[h]
    \centering
    \includegraphics[width=0.45\textwidth]{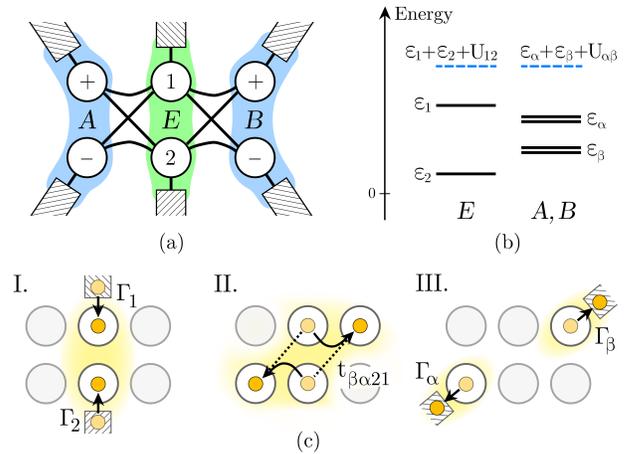}
    \captionsetup{justification=justified,singlelinecheck=false}
    \caption{(a)~Schematics of the setup, with entangler ($E$) and detector ($A$ and $B$) subsystems, quantum dots (circles), electronic leads (hatched rectangles), and key tunneling couplings (thick lines) shown. (b) A typical configuration of $E$ and $A, B$ dot energies $\epsilon_1$, $\epsilon_2$, $\epsilon_\alpha$, $\epsilon_\beta$ with $\alpha = A\pm$, $\beta = B\pm$ and inter-dot charging energies $U_{12}$, $U_{\alpha \beta}$ shown, fulfilling the two-particle resonance condition $\epsilon_1+\epsilon_2+U_{12}=\epsilon_\alpha+\epsilon_\beta + U_{\alpha \beta}$. (c)~The three tunneling processes in the pair transfer cycle: (I)~Entangler dots $1$, $2$ populated with rates $\Gamma_1$, $\Gamma_2$. (II)~Coherent cotunneling of electrons from entangler to detector dots (here $\alpha = A-$ and $\beta = B+$), with amplitude $t_{\beta \alpha 2 1}$. (III) Electrons in $\alpha$, $\beta$ emitted to detector leads, with rates $\Gamma_\alpha$, $\Gamma_\beta$.}
    \label{QDsystem}
\end{figure}

Here we propose such an entanglement scheme, based on coherent electron cotunneling \cite{Averin&Nazarov} in a quantum dot system, see Fig.~\ref{QDsystem}. During the cotunneling event, of the order of picoseconds \cite{PhysRevB.78.155309, NewJPhys.14.083003}, the electrons are entangled via two-particle interference \cite{PhysRevLett.92.026805,PhysRevLett.102.106804} and simultaneously transferred to the detectors, fully preserving coherence \cite{PhysRevLett.96.036804, doi:10.1021/nl801689t}. We show, based on the full transfer statistics \cite{PhysRevLett.94.210601,PhysRevB.65.075317}, that the entanglement can conveniently be detected by violating a Bell inequality (BI) formulated in terms of low-frequency current cross-correlators \cite{PhysRevB.66.161320, PhysRevLett.92.026805, PhysRevLett.91.147901}. Moreover, for long enough detector dot lifetimes, measurements of coincident electrons in the detector dots can be performed with real-time charge counting techniques \cite{PhysRevB.79.035314}. This opens up for an unambiguous BI-test of orbitally entangled electrons in nanosystems based on short-time measurements.

The combined entangler-detector system, shown in Fig.~\ref{QDsystem}, consists of two entangler ($E$) dots, $1$ and $2$, and four detector dots, $\alpha=A\pm$ at $A$ and $\beta=B\pm$ at $B$. Each dot $\gamma=1,2,\alpha,\beta$ has one active, spin-degenerate level at energy $\epsilon_{\gamma}$ and is tunnel-coupled to an electronic lead, with a rate $\Gamma_{\gamma}$. The six dots are further coupled to each other (nearest neighbors) with tunneling amplitudes $t_{\alpha 1},t_{\alpha 2},t_{\beta 1},t_{\beta 2},t_{12},t_{A+A-}$ and $t_{B+B-}$. The Coulomb interaction energy between different dots, $\gamma \neq \gamma'$, is $U_{\gamma\gamma'}$. Due to strong on-site repulsion, double occupation of the dots is prevented. Consequently, the spin degree of freedom only leads to a renormalization of tunneling rates and is hereafter neglected.

As illustrated in Fig. \ref{QDsystem}, the dot level energies $\epsilon_{\gamma}$ are tuned to suppress single-particle tunneling between entangler and detector dots (as well as between the two entangler dots). In addition, to make cotunneling the dominating entangler-detector transport mechanism, the energies are chosen to optimize the conditions $\epsilon_\alpha + \epsilon_\beta + U_{\alpha \beta} \approx \epsilon_1 + \epsilon_2 + U_{12}$ for resonant two-particle tunneling between entangler dots 1,2 and detector dots $\alpha,\beta$. The amplitude for the cotunneling is denoted $t_{\beta\alpha21}$. Moreover, cotunneling between the entangler dots and two dots at the same detector, $A$ or $B$, is tuned off resonance.

The dot-lead tunneling rates obey $\Gamma_{\alpha},\Gamma_{\beta} \gg \Gamma_1,\Gamma_2 \gg t_{\beta\alpha21}/\hbar$. Hence, the cotunneling is much slower than the dot-lead tunneling (sequential) and back-tunneling from the detector to the entangler dots is suppressed. Moreover, in this regime there are at most two particles in the dot-system at the same time. Furthermore, taking $\hbar\Gamma_{\alpha} \gg t_{A+A-}$ and $\hbar\Gamma_{\beta} \gg t_{B+B-}$, tunneling between the detector dots can be neglected. The two entangler leads are kept at a finite bias while the four detector dots are grounded. Throughout the paper we consider the high-bias regime; the energy difference between dot system states differing by one electron is well inside the bias window and the lead temperature can effectively be put to zero.

\begin{figure}[h]
    \centering
    \includegraphics[width=0.35\textwidth]{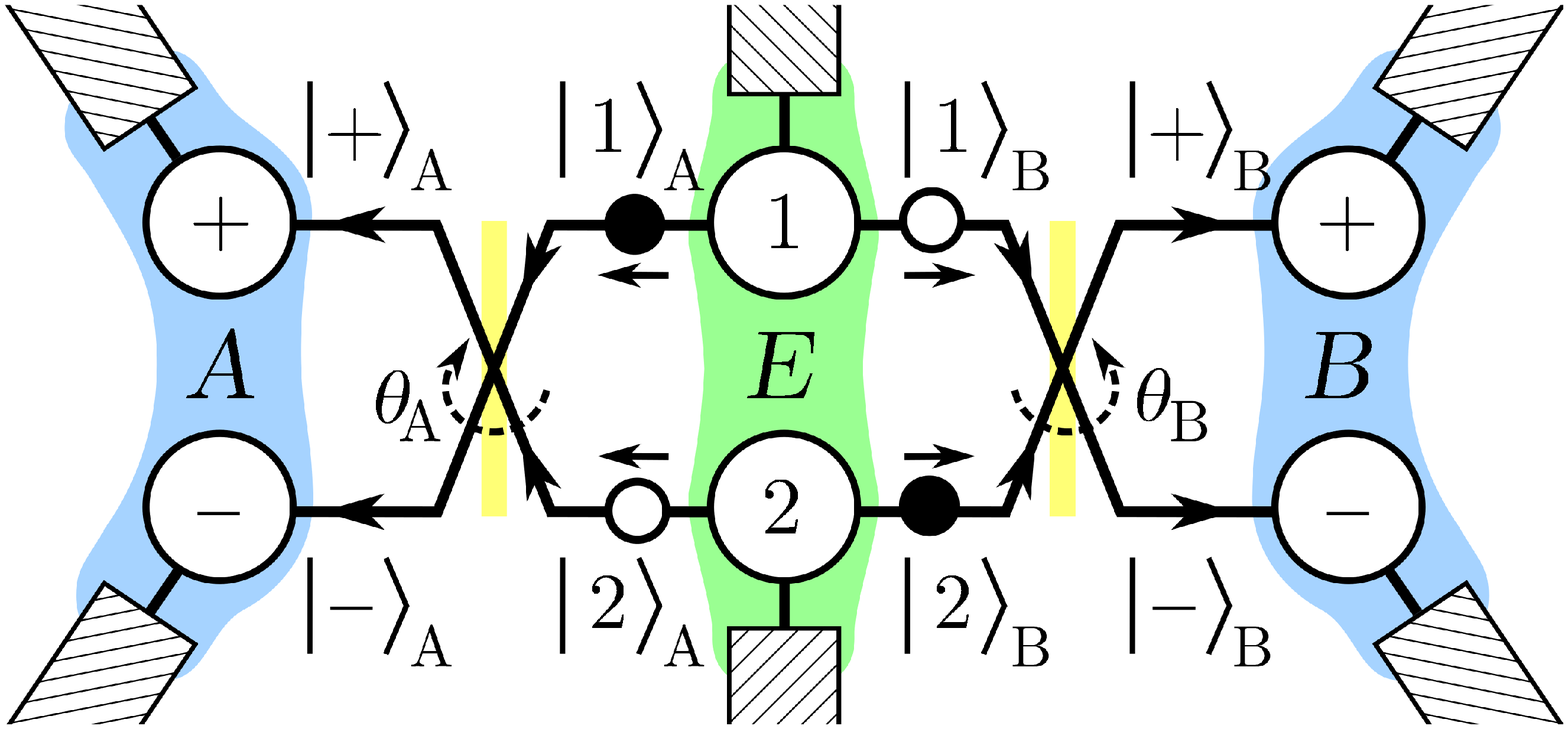}
    \captionsetup{justification=justified,singlelinecheck=false}
    \caption{Schematics of the entanglement creation and state rotation during the cotunneling, emphasizing the orbital states $|1\rangle_A|2\rangle_B$ ($|2 \rangle_A|1\rangle_B$) for electrons emitted towards $A$ ($B$). The state $|\psi\rangle_{\text{ent}}$ in Eq.~\ref{Eq:Emitted state} is a coherent two-particle superposition, describing the entangler dot electrons being emitted $1 \rightarrow A$, $2 \rightarrow B$ (filled dots, state $|1\rangle_A|2\rangle_B$) or $2 \rightarrow A$, $1 \rightarrow B$ (empty dots, $|2\rangle_A|1\rangle_B$). The tunability of the tunneling amplitudes $t_{\alpha 1}$ etc., here illustrated as beam splitters (shaded yellow) parameterized by angles $\theta_A$, $\theta_B$, allows for a rotation in orbital $\{ |1\rangle$, $|2\rangle$ $\rightarrow$ $|+\rangle$, $|-\rangle\}$-space of $A$ and $B$.}
    \label{Orbital entanglement}
\end{figure}

We first give a physically compelling picture of the ideal working of the entangler-detector system, a detailed analysis follows below. In Fig.~\ref{QDsystem}, the tunneling processes in the pair transfer cycle, from entangler to detector leads, are shown. (I) Starting from an empty system, electrons tunnel sequentially into the two entangler dots 1 and 2. (II) The electrons in the entangler dots cotunnel resonantly to the detector dots, one electron to $\alpha$ and one to $\beta$. This process is coherent and key to the entangler-detector scheme in the following ways: (1) The electrons can cotunnel in two ways, from $1$ to $\alpha$ and $2$ to $\beta$ or $1$ to $\beta$ and $2$ to $\alpha$. This gives rise to orbital entanglement via two-particle interference as illustrated in Fig.~\ref{Orbital entanglement}. The quantum state emitted from the entangler can be written
\begin{equation}
	|\psi \rangle_{\text{ent}} = c_{21} |2\rangle_A|1\rangle_B-c_{12}|1\rangle_A|2\rangle_B
	\label{Eq:Emitted state}
\end{equation}
with $|1\rangle_A$ denoting an electron emitted from dot $1$ towards detector $A$ etc. and $c_{12}$, $c_{21}$ constants depending on the system properties. We note that for $|c_{12}| = |c_{21}| = \frac{1}{\sqrt{2}}$, $|\psi\rangle_{\text{ent}}$ is maximally entangled. (2)~By tuning, with electrostatic gates, the dot-entangler tunneling amplitudes $t_{1\alpha}$ etc., the emitted state $|\psi\rangle_{\text{ent}}$ can be locally rotated in the $|1\rangle, |2\rangle$ basis during the transfer to the detector dots, see Fig.~\ref{Orbital entanglement}.

(III) At the detector dots, the two particles tunnel sequentially out to the leads. As will be described below, depending on the values of $\Gamma_\alpha$, $\Gamma_\beta$ the entanglement can be detected via violation of a BI formulated in terms of either low-frequency current cross-correlations or short-time measurements of joint detection probabilities.

To substantiate the qualitative picture above we first recall that uncertainty principle arguments give a cotunneling time $\sim \hbar/\Delta E$, where $\Delta E$ is determined by the energy of the classically forbidden state, virtually populated during the tunneling \cite{Averin&Nazarov}. Here we take the cotunneling time, typically of order of picoseconds in quantum dot systems \cite{PhysRevB.78.155309}, much smaller than the decoherence time $1/\Gamma_{\varphi}$, of the order of nanoseconds \cite{PhysRevLett.91.226804, PhysRevLett.105.246804}. The cotunneling process is thus coherent, as recently demonstrated for double-dot systems \cite{PhysRevLett.96.036804}, and can be treated within an effective Hamiltonian approach.

To this end, we write the total Hamiltonian as $H = H_0 + V$. The term $H_0$ describes the single-particle dot levels, the Coulomb repulsion between the dots, the electronic leads and the dot-lead tunneling. The perturbation $V$ describes the dot-dot tunneling. The cotunneling dynamics is obtained from a Schrieffer-Wolff transformation \cite{PhysRev.149.491}, where $V$ is eliminated to leading order, see \cite{SI}. Besides cotunneling processes, the transformation yields renormalization terms which are absorbed into the dot-level energies, the Coulomb interaction strengths and the dot-lead couplings in $H_0$. Furthermore, neglecting terms describing processes much slower than competing sequential dot-lead tunneling or resonant two-particle cotunneling, we obtain the effective cotunneling Hamiltonian
\begin{equation}
H_{\mathrm{eff}} = H_0 + \sum\limits_{\alpha \beta} \left( t_{\beta \alpha 2 1}d^\dagger_{\beta} d^\dagger_{\alpha} d_{2} d_{1} + H.c. \right)
\label{Eq:Effective Hamiltonian}
\end{equation}
with $d^\dagger_\beta$ ($d_\beta$) the creation (annihilation) operator of an electron in dot $\beta$. Here the two-particle tunneling amplitudes $t_{\beta \alpha 2 1}$ are related to the single-particle amplitudes through
\begin{equation}
t_{\beta \alpha 2 1} = \frac{t_{\beta 1}t_{\alpha 2}}{\Delta E_{\beta \alpha}} - \frac{t_{\alpha 1}t_{\beta 2}}{\Delta E_{\alpha \beta}}
\label{Eq:Co-tunneling amplitude}
\end{equation}
with the energies $\Delta E_{\beta\alpha} = \Delta E_{\alpha \beta}[\alpha \leftrightarrow \beta]$ depending on $\epsilon_\gamma$ and $U_{\gamma \gamma'}$, see \cite{SI}. In particular, at two-particle resonance, $\epsilon_\alpha + \epsilon_\beta +U_{\alpha \beta} = \epsilon_1+\epsilon_2+U_{12} \equiv E_r$, $1/\Delta E_{\beta \alpha} = (E_r - [\epsilon_2+\epsilon_\beta + U_{2\beta}])^{-1} +(E_r-[\epsilon_1+\epsilon_\alpha+U_{1\alpha}])^{-1}$. 

Importantly, the terms in Eq.~\eqref{Eq:Co-tunneling amplitude} correspond to two different two-particle paths, $1\rightarrow\beta$, $2\rightarrow \alpha$ and $1\rightarrow \alpha$, $2\rightarrow\beta$ (see Fig.~\ref{Orbital entanglement}), that take the two electrons between the entangler and detector dots, with tunneling times $\hbar / \Delta E_{\beta \alpha}$ and $\hbar/\Delta E_{\alpha \beta}$, respectively. Hence, the total cotunneling amplitude $t_{\beta \alpha 2 1}$ is a coherent superposition of the individual two-particle tunneling amplitudes. The minus sign between the terms is a consequence of fermionic exchange. Note that the amplitude for both electrons to tunnel to the same detector side, $A$ or $B$, is negligible since the process is off-resonant (due to level detuning and Coulomb interaction $U_{A+A-}, U_{B+B-}$).

Starting from the state with both entangler dots occupied, time evolution governed by $H_\text{eff}$ gives that the two-particle state tunneling out into the detector dots is $|\psi\rangle_\text{dot} \propto \sum_{\alpha \beta} t_{\beta \alpha 21}|\alpha\rangle|\beta\rangle$, with $|\alpha\rangle = |+\rangle_A,|-\rangle_A$ and $|\beta\rangle = |+\rangle_B,|-\rangle_B$. To arrive at the result in Eq. (\ref{Eq:Emitted state}), we focus on the experimentally relevant regime where the specific dot ($\pm$ in $\alpha,\beta$) dependent part of the Coulomb interactions between detector dots at $A$ and $B$, $U_{\alpha \beta}$, as well as between entangler and detector dots, $U_{1\alpha}$ etc., is negligibly small compared to the relevant energy level differences in $\Delta E_{\alpha \beta}, \Delta E_{\beta \alpha}$. At two-particle resonance, this regime implies $\epsilon_\alpha \approx \epsilon_A$, $\epsilon_\beta \approx \epsilon_B$ and thus $\Delta E_{\alpha\beta}=\Delta E_{AB}, \Delta E_{\beta\alpha}=\Delta E_{BA}$, independent of $\alpha,\beta$ (see Fig. \ref{QDsystem}). In addition, we consider the single-particle tunneling amplitudes to be tuned to fulfill the relation $|t_{A+1}/t_{A-1}|=|t_{A-2}/t_{A+2}|$. Under these conditions, with Eq. (\ref{Eq:Co-tunneling amplitude}), one can write the state in the detector dots
\begin{equation}
|\psi \rangle_\text{dot} = (S_A\otimes S_B)|\psi\rangle_\text{ent}, \hspace{0.3cm} S_{i} = \left( \begin{array}{cc}\sin\theta_{i}& \cos\theta_{i} \\ \cos\theta_{i}& -\sin\theta_{i} \end{array} \right)
\label{Transformation}
\end{equation}
with $S_i$ the scattering matrix for the effective beam splitter at detector $i=A,B$, see Fig.~\ref{Orbital entanglement}. Here we have introduced the parameterization of the tunneling amplitudes $t_{A+1} = t_{A1}\cos\theta_A$, $t_{A-1} = t_{A1}\sin\theta_A$ etc., with $|t_{A1}|^2 = |t_{A+1}|^2+|t_{A-1}|^2$ etc. Moreover, we can directly identify $c_{12} = t_{A1}t_{B2}/\Delta E_{AB}$ and $c_{21} = t_{A2}t_{B1}/\Delta E_{BA}$ in Eq.~\eqref{Eq:Emitted state}. Importantly, Eq.~\eqref{Transformation} verifies the claim that the cotunneling from the entangler to the detector dots can be described as an emission of an entangled state, $|\psi\rangle_\text{ent}$, locally rotated in orbital space before arriving to the detector dots.

Detection of the entanglement of $|\psi\rangle_{\text{ent}}$ can be performed by transport measurements. In the high-bias regime considered, the full transport statistics of the entangler-detector system can be described within the framework of a Markovian quantum master equation $\frac{d\rho}{dt} = \mathcal{L}_\chi[\rho]$ for the reduced density operator $\rho$, with the Liouvillian superoperator \cite{breuer2007theory,EurophysLett.69.3, RevModPhys.81.1665}
\begin{eqnarray}
  \mathcal{L}_\chi [\rho] = -\frac{i}{\hbar} \lbrack H_{ S}, \rho
  \rbrack + \sum\limits_{\gamma} \bigg[ \Gamma_{\gamma} \Big( f_{
    \gamma} \mathcal{D}_{-\chi_\gamma}[d^\dagger_{\gamma}, \rho] \nonumber \\
  +(1-f_{ \gamma})\mathcal{D}_{\chi_\gamma}[d_{\gamma},\rho]\Big) +
  \frac{\Gamma_{ \varphi}}{2} \mathcal{D}_{0}[d^\dagger_{\gamma}
  d_{\gamma},\rho] \bigg]
\label{Master equation}
\end{eqnarray}
where $H_S$ is the Hamiltonian of the dot system, $f_\gamma$ is the lead Fermi function with $f_\gamma = 1$ (0) for $\gamma = 1,2$ ($\alpha,\beta$), $\mathcal{D}_\chi[L,\rho]= e^{ i\chi}L \rho L^\dagger - \frac{1}{2} \left\{ L^\dagger L,\rho \right\}$ is the dissipator, and $\chi_\gamma$ is a counting field for lead $\gamma$. The last term in Eq.~\eqref{Master equation} describes dephasing with a rate $\Gamma_\varphi$ independent of $\gamma$.

The cumulant generating function $F_\chi$ is obtained as the eigenvalue of $\mathcal{L}_\chi$ fulfilling the condition $\lim_{\chi \to 0}F_\chi\to0$. To leading order in the tunneling amplitudes $t_{\beta \alpha 2 1}$ we find
\begin{equation}
F_\chi = \sum \limits_{\alpha \beta}\left(e^{i(\chi_{\alpha}+\chi_{\beta}-\chi_{1}-\chi_{2})}-1\right)P_{\alpha \beta}
\label{Eq:Cumulant generating function}
\end{equation}
where the transfer rates $P_{\alpha \beta}$ are given by
\begin{equation}
P_{\alpha \beta} = \frac{|t_{\beta \alpha 2 1}|^2\left(\Gamma_\alpha+\Gamma_\beta+\Gamma_\varphi\right)}{\frac{\hbar^2}{4}\left(\Gamma_\alpha+\Gamma_\beta+\Gamma_\varphi\right)^2+\epsilon_{12\alpha \beta}^2}
\label{Eq:Co-tunneling rates}
\end{equation}
with $\epsilon_{12\alpha \beta} \equiv \epsilon_\alpha + \epsilon_\beta + U_{\alpha \beta} - [ \epsilon_1 + \epsilon_2 + U_{12}]$ the energy away from two-particle resonance. The generating function $F_\chi$ in Eq.~\eqref{Eq:Cumulant generating function}, a sum of terms $P_{\alpha \beta}(e^{i(\chi_{\alpha}+\chi_{\beta}-\chi_{1}-\chi_{2})}-1)$, gives a physically clear picture of the full transport statistics: There are four independent, elementary events of electron pair transfer from the entangler leads $1,2$ to the detector leads $\alpha$ and $\beta$ with rates $P_{\alpha \beta}/\hbar$. 

To formulate a BI-test in terms of transport quantities, we consider either the dephasing-broadened regime, where $\Gamma_\varphi \gg \Gamma_\alpha,\Gamma_\beta$, or the situation where $\Gamma_\alpha,\Gamma_\beta$ are independent of the specific dot ($\pm$ in $\alpha,\beta$). In both cases $P_{\alpha \beta} \propto |t_{\beta \alpha 2 1}|^2$, with a proportionality constant independent of $\alpha,\beta$. This allows us to identify, directly from
Eq.~\eqref{Eq:Cumulant generating function}, $P_{\alpha \beta}/\sum_{\alpha \beta}P_{\alpha \beta}$ as the probability that a pair of electrons emitted from $E$ is jointly detected, with one electron in lead $\alpha$ and one in $\beta$. Based on these joint detection probabilities, we can directly formulate a CHSH BI \cite{PhysRevLett.91.157002,PhysRevLett.23.880} as
\begin{equation}
S = |E_{AB}- E_{A'B}+E_{AB'}+E_{A'B'}| \leq 2,
\label{Eq:Bell parameter}
\end{equation}
with the correlation functions $E_{AB} =
\frac{P_{A+B+}-P_{A+B-}-P_{A-B+}+P_{A-B-}}{P_{A+B+}+P_{A+B-}+P_{A-B+}+P_{A-B-}}$ and where $A, A', B, B'$ denote different detector settings, i.e., here different single-particle tunneling amplitudes. Importantly, the rates $P_{\alpha \beta}$ can be obtained from the current cross-correlation $S_{\alpha \beta} = e^2\frac{\partial^2F_\chi}{\partial i\chi_\alpha \partial i\chi_\beta}|_{\chi = 0} = e^2 P_{\alpha \beta}$, allowing for an experimental test of entanglement via a violation of the BI. We note that for the state $|\psi\rangle_{\text{ent}}$, the maximal Bell parameter is $S_{\text{max}} = 2\sqrt{1+\sin^2\theta}$ \cite{PhysRevLett.91.157002}, where $\theta =2\arctan(c_{12}/c_{21})$, i.e., all entangled states can in principle be detected by a BI violation.

Beyond being insensitive to decoherence, does our setup offer any additional advantages, not present in existing proposals? The answer is yes. By tuning the detector tunneling barriers such that the rates $\Gamma_\alpha$, $\Gamma_\beta$ become small, typically in the sub-MHz range, the tunneling on and off the detector dots can be monitored in real time, via, e.g., time-dependent electrical currents flowing through quantum point contacts capacitively coupled to the dots (not shown in Fig.~\ref{QDsystem}) \cite{PhysRevLett.96.076605, PhysRevB.74.195305, ProcNatlAcadSci.106.10116}. In particular, by simultaneously detecting the charge transfers at dots $\alpha$ and $\beta$, as demonstrated in Ref.~\cite{PhysRevB.79.035314}, one can identify the coincidence probabilities for individual, emitted pairs arriving at the detector dots or leads. This opens up for an entanglement test based on a violation of a BI formulated in terms of short-time joint detection probabilities, in direct analogy to quantum optics.

Theoretically, the short-time properties of the correlated charge transfer can be described by the electronic analogue \cite{PhysRevB.85.165417} of Glauber's second degree of coherence \cite{PhysRev.130.2529} in quantum optics,
\begin{equation}
g^{(2)}_{\alpha \beta} (\tau) = \frac{ \langle\langle J_{\alpha} \Omega (\tau) J_{\beta} \rangle\rangle + \langle\langle J_{\beta} \Omega (\tau) J_{\alpha} \rangle\rangle}{2\langle\langle J_{\alpha} \rangle\rangle\langle\langle J_{\beta} \rangle\rangle}
\end{equation}
where $\langle \langle A \rangle \rangle$ is the stationary expectation value of operator $A$, $\Omega (\tau) = e^{\mathcal{L}_0\tau}$ is the master equation propagator and $J_\gamma = \partial_{ i\chi_\gamma}\mathcal{L}_\chi \big|_{\chi = 0}$ is the current superoperator at lead $\gamma = \alpha,\beta$. We note that the Liouvillian $\mathcal{L}_\chi$ is given by Eq.~\eqref{Master equation} and that, by definition, $e\langle\langle J_\alpha \rangle\rangle =I_\alpha$. To leading order in $t_{\beta \alpha 2 1}$, we obtain
\begin{equation}
g^{(2)}_{\alpha \beta}(\tau) = P_{\alpha \beta}\frac{\Gamma_{\alpha}\Gamma_{\beta}}{\Gamma_{\alpha} + \Gamma_{\beta}} \frac{e^2}{2I_{\alpha}I_{\beta}}(e^{-\Gamma_\alpha \tau}+e^{-\Gamma_\beta \tau})
\end{equation}

Several important conclusions can be drawn from this result. First, $g^{(2)}_{\alpha \beta}(\tau)\gg 1$ for all times $\tau \lesssim 1/\Gamma_\alpha, 1/\Gamma_\beta$ (terms in $g^{(2)}_{\alpha \beta}(\tau)$ of order unity are neglected), a typical signature for pair transport \cite{Friberg1985311}. Second, the decay timescales of the correlations, $1/\Gamma_\alpha, 1/\Gamma_\beta$ are set by the tunneling times out of the dots $\alpha$ and $\beta$. Third, $g^{(2)}_{\alpha \beta} (\tau)$ by construction describes the correlations between electrons arriving at the leads $\alpha,\beta$, i.e., tunneling out of the corresponding dots $\alpha, \beta$. To obtain the probabilities that two electrons arrive coincidentally at the detector dots $\alpha$ and $\beta$, we have to integrate $g_{\alpha \beta}^{(2)}(\tau)$ over all times, i.e., accounting for all possible emissions from the dots to the leads. This gives the quantity (not normalized with currents)
\begin{equation}
2I_{\alpha}I_{\beta}\int \limits^{\infty}_{0} d\tau g^{(2)}_{\alpha \beta}(\tau)= e^2 P_{\alpha \beta}
\end{equation}
which thus gives the same probabilities $P_{\alpha \beta}/\sum_{\alpha \beta}P_{\alpha \beta}$ as in the long-time limit, Eqs.~\eqref{Eq:Cumulant generating function} and \eqref{Eq:Co-tunneling rates}, producing a consistent theoretical picture of the charge transfer.

Interestingly, in contrast to charge detection via long-time current correlations, real-time detection of coincident probabilities has typically non-ideal efficiencies, $\eta_\alpha,\eta_\beta < 1$. The efficiency at $A$ is given by $\eta_\alpha = e^{-\Gamma_\alpha/\Delta\omega}$ (and similarly at $B$), where $\Delta\omega$ is the bandwith of the detector. Based on existing experiments, near-unity efficiency of the detection requires $\Gamma_{\alpha},\Gamma_{\beta}$ well below 1 MHz \cite{PhysRevLett.96.076605, PhysRevB.74.195305, ProcNatlAcadSci.106.10116}.

In conclusion, we have proposed a quantum dot based entangler-detector system that generates and detects orbitally entangled electrons on a timescale much shorter than the decoherence time. The main idea is to use cotunneling for both the entangling and state manipulation during the transfer to the detectors. Recent experimental demonstrations of highly controllable multiple quantum dots systems \cite{ApplPhysLett.101.10,ApplPhysLett.104.11, ApplPhysLett.104.183111} together with existing short-time charge detection techniques \cite{PhysRevLett.96.076605, PhysRevB.74.195305, ProcNatlAcadSci.106.10116} make the realization of our scheme within experimental reach.

\bibliographystyle{apsrev4-1}
\bibliography{sources}

\end{document}

\title{Supplemental Material for \\ Sub Decoherence Time Generation and Detection of Orbital Entanglement}
\author{F. Brange}
\author{O. Malkoc}
\author{P. Samuelsson}
\affiliation{Department of Physics, Lund University, Box 118, SE-221 00 Lund, Sweden}

\begin{abstract}
\end{abstract}

\maketitle

Here we derive the effective cotunneling Hamiltonian in Eq.~(2) in the main text. We start by considering the isolated dot-system described by the Hamiltonian $H_S = H_0 + V$, where $H_0 = \sum \limits_{\gamma} \epsilon_\gamma n_\gamma + \sum \limits_{\gamma \gamma '} \frac{1}{2}U_{\gamma \gamma '} n_\gamma n_{\gamma '}$, with single-particle dot-levels $\epsilon_\gamma$ and the Coulomb repulsion $U_{\gamma \gamma '}$ between different dots ($\gamma \neq \gamma'$). Here $n_\gamma = d^\dagger_\gamma d_{\gamma}$ is the number operator. The term $V = \sum \limits_{\gamma \gamma '} \frac{1}{2} (t_{\gamma \gamma '} d^\dagger_\gamma d_{\gamma '} + t^*_{\gamma \gamma '} d^\dagger_{\gamma '} d_{\gamma})$ describes the tunneling, with amplitudes $t_{\gamma \gamma '}$, between nearest neighbor dots. We set the tunneling amplitudes between detector dots effectively to zero as tunneling out to the leads will dominate over back-tunneling to other dots. For the remaining single-particle dot-dot tunneling couplings we assume off resonance conditions, allowing us to eliminate $V$ to first order in $H_S$ by means of a generalized Schrieffer-Wolff transformation $H_S \rightarrow e^{-S}H_Se^S \equiv H_{\text{eff}}$ \cite{PhysRev.149.491}.

The generator $S$ of the transformation is obtained from the condition \cite{PhysRev.149.491} 
\begin{equation}
[S,H_0] = -V
\label{S generator condition}
\end{equation}
To find an explicit expression of $S$, we first introduce an operator
\begin{equation*}
A_{\gamma \gamma'} \! =  \! \sum \limits_{B} \! \left( \! \frac{1}{\epsilon_{\gamma} \! - \!  \epsilon_{\gamma  '}\! +\!\! \sum \limits_{\delta' \in B} \! (U_{\gamma \delta'}\! -\! U_{\gamma  ' \delta'})} \! \prod \limits_{\delta \in B} \! n_\delta \! \prod \limits_{\bar{\delta} \in \bar{B}}\! (1\! -\! n_ {\bar{\delta}}) \! \right)
\label{A operator}
\end{equation*}
for $\gamma \neq \gamma'$. To clarify the notation, we point out that $A_{\gamma \gamma'}$ contains number operators acting only on the four dots (of the in total six) which are not $\gamma$, $\gamma'$. Out of these four dots, $B$ denotes a configuration of $n = 0-4$ occupied dots and $\bar{B}$ denotes the remaining $4-n$ unoccupied dots, i.e., the complement of $B$. The sum in $A_{\gamma \gamma'}$ runs over all 16 possible configurations $B$. Moreover, each term in the sum contains a denominator $\epsilon_{\gamma} \! - \!  \epsilon_{\gamma  '}\! +\! \sum \limits_{\delta' \in B} (U_{\gamma \delta'}\! -\! U_{\gamma  ' \delta'})$, the energy difference associated with single-particle tunneling between $\gamma'$ and $\gamma$ in presence of the occupation configuration $B$ of the four other dots.

In terms of $A_{\gamma \gamma'}$, the generator $S$ can then be expressed as
\begin{equation}
S = \sum \limits_{\gamma \gamma '} \frac{1}{2}A_{\gamma \gamma '}(t_{\gamma \gamma '} d^\dagger_{\gamma}d_{\gamma '} - t^*_{\gamma \gamma '} d^\dagger_{\gamma '}d_{\gamma})
\label{Generator}
\end{equation}
To prove this, we note that $A_{\gamma \gamma '}$ is Hermitian and fulfills the property $A_{\gamma \gamma '} = -A_{\gamma ' \gamma}$. Furthermore, we note that $[A_{\gamma \gamma '}, H_0] = 0$, since both operators only consist of number operators. Importantly, we have
\begin{multline*}
[A_{\gamma \gamma '}d^\dagger_{\gamma}d_{\gamma '},H_0] = A_{\gamma \gamma '}[d^\dagger_{\gamma}d_{\gamma  '},H_0] \\
= A_{\gamma \gamma '} \left[ \epsilon_{\gamma  '}-\epsilon_{\gamma}+\sum \limits_{\delta'} (U_{\gamma ' \delta'} - U_{\gamma \delta'}) n_{\delta'} \right] d^\dagger_{\gamma}d_{\gamma  '} \\
= -\sum \limits_{B}\left[ \prod \limits_{\delta \in B} n_\delta \prod \limits_{\bar{\delta} \in \bar{B}} (1- n_{\bar{\delta}})\right] d^\dagger_{\gamma}d_{\gamma '} = - d^\dagger_{\gamma}d_{\gamma '}
\end{multline*}
where we note that the sum on the second line is over all dots except $\gamma$ and $\gamma'$. We then get
\begin{multline*}
[S,H_0] = \left[ \sum \limits_{\gamma \gamma '} \frac{1}{2}A_{\gamma \gamma '}(t_{\gamma \gamma '} d^\dagger_{\gamma}d_{\gamma '} - t^*_{\gamma \gamma '} d^\dagger_{\gamma '}d_{\gamma}), H_0\right] \\ = \sum \limits_{\gamma \gamma '} \frac{1}{2}\left(t_{\gamma \gamma '} [A_{\gamma \gamma '}d^\dagger_{\gamma}d_{\gamma '},H_0] +t^*_{\gamma \gamma '}[A_{\gamma' \gamma }d^\dagger_{\gamma'}d_{\gamma},H_0] \right) \\
= -\sum \limits_{\gamma \gamma '} \frac{1}{2}\left(t_{\gamma \gamma '} d^\dagger_{\gamma}d_{\gamma '}+t^*_{\gamma \gamma '}d^\dagger_{\gamma '}d_{\gamma}\right) = - V
\end{multline*}
which shows that $S$ in Eq.~\eqref{Generator} fulfills Eq.~\eqref{S generator condition}.

Next, we compute the effective Hamiltonian to leading order in $V$ from \cite{PhysRev.149.491}
\begin{equation}
H_{\mathrm{eff}} = H_0 + \frac{1}{2}[S,V]
\end{equation}
To evaluate the commutator of $S$ and $V$, we first consider the commutator of a single term of $S$, $A_{\delta \delta '}(t_{\delta \delta '} d^\dagger_{\delta} d_{\delta '} - t^*_{\delta \delta '} d^\dagger_{\delta '}d_{\delta})$, and a single term of $V$, $t_{\gamma \gamma '} d^\dagger_{\gamma} d_{\gamma '} + t^*_{\gamma \gamma '}d^\dagger_{\gamma'} d_{\gamma} $.

Four different terms are produced by the commutator, dependent on the relation between $\gamma$ and $\delta$ and between $\gamma'$ and $\delta'$ (note that $\gamma \neq \gamma'$ and $\delta \neq \delta'$). Below we evaluate and discuss them one by one, making use of the assumption that there can be maximally two electrons in the dot system at the same time.

\noindent \emph{(i)} $\delta = \gamma, \delta' = \gamma'$

We have the commutator
\begin{multline}
\frac{1}{2}\lbrack A_{\gamma \gamma '}(t_{\gamma \gamma '} d^\dagger_{\gamma} d_{\gamma '} - t^*_{\gamma \gamma '} d^\dagger_{\gamma '}d_{\gamma}), t_{\gamma \gamma '} d^\dagger_{\gamma} d_{\gamma '} +  t^*_{\gamma \gamma '} d^\dagger_{\gamma '} d_{\gamma} \rbrack \\
= |t_{\gamma \gamma '}|^2 A_{\gamma \gamma '}(n_{\gamma}-n_{\gamma '})
\end{multline}
This term only consists of number operators and can therefore be eliminated by means of a renormalization of the single-particle energy levels $\epsilon_\gamma$ and the interaction strengths $U_{\gamma \gamma '}$.

\newpage
\noindent \emph{(ii)} $\delta = \gamma, \delta' \neq \gamma'$ $\emph{and}$ \emph{(iii)} $\delta \neq \gamma, \delta' = \gamma'$

For $\delta = \gamma, \delta' \neq \gamma'$, we have the commutator
\begin{multline}
\frac{1}{2}\lbrack  A_{\gamma \delta '}(t_{\gamma \delta '} d^\dagger_{\gamma} d_{\delta '} - t^*_{\gamma \delta '} d^\dagger_{\delta '}d_{\gamma}),t_{\gamma \gamma '} d^\dagger_{\gamma} d_{\gamma ' } + t^*_{\gamma \gamma '} d^\dagger_{\gamma '} d_{\gamma} \rbrack \\
= \! -\frac{1}{2}A_{\gamma \delta '}[n_{\gamma '} \! \rightarrow \! n_{\gamma}] (t_{\gamma \delta '}t^*_{\gamma \gamma '} d^\dagger_{\gamma '} d_{\delta '}\! +\! t^*_{\gamma \delta '}t_{\gamma \gamma '} d^\dagger_{\delta '} d_{\gamma '})
\label{Single-particle tunneling}
\end{multline}
where $A_{\gamma \delta '}[n_{\gamma '} \! \rightarrow \! n_{\gamma}]$  denotes $A_{\gamma \delta '}$, with $n_{\gamma '}$ replaced by $n_{\gamma}$.  This term describes tunneling of a single particle, first from $\delta'$ to $\gamma$ and then from $\gamma$ to $\gamma'$, or vice versa. A similar term is obtained when $\delta \neq \gamma$, $\delta ' = \gamma '$.

However, in the considered parameter regime (specified in the main text), these tunneling processes are not important for the dynamics of the dot system, as there is always a competing process which dominates. To be explicit, there are three different processes. First, if an electron occupies one of the detector dots, the most likely process is tunneling to the lead and not tunneling to another dot, since the dot-lead couplings are much stronger than the dot-dot couplings. Second, if one of the entangler dots is occupied, the most likely process is that the second entangler dot becomes occupied. Third, when both entangler dots are occupied, the dominating process is cotunneling, see (iv), since this process is close to resonance in contrast to single-particle tunneling. The single-particle tunneling processes described by Eq.~\eqref{Single-particle tunneling} are thus neglected.

\noindent \emph{(iv)} $\delta \neq \gamma, \delta' \neq \gamma'$

We have the commutator
\begin{multline}
\frac{1}{2}\lbrack A_{\delta \delta '}(t_{\delta \delta '} d^\dagger_{\delta} d_{\delta '} - t^*_{\delta \delta '} d^\dagger_{\delta '}d_{\delta}),t_{\gamma \gamma '} d^\dagger_{\gamma} d_{\gamma '} + t^*_{\gamma \gamma '} d^\dagger_{\gamma '} d_{\gamma} \rbrack \\
= \frac{1}{2} \left( \frac{1}{\epsilon_{\delta}+U_{\gamma \delta} - \epsilon_{\delta '}-U_{\gamma \delta '}} - \frac{1}{\epsilon_{\delta}+U_{\gamma ' \delta} - \epsilon_{\delta '}-U_{\gamma' \delta '}} \right) \\
\times\bigg( t_{\delta \delta '}t_{\gamma \gamma '} d^\dagger_{\delta}d^\dagger_{\gamma}d_{\gamma '}d_{\delta '} +t^*_{\delta \delta '}t^*_{\gamma \gamma '} d^\dagger_{\delta '}d^\dagger_{\gamma '} d_{\gamma}d_{\delta}\\
+t^*_{\delta \delta '}t_{\gamma \gamma '} d^\dagger_{\delta '}d^\dagger_{\gamma}d_{\gamma '}d_{\delta}+t_{\delta \delta '}t^*_{\gamma \gamma '} d^\dagger_{\delta}d^\dagger_{\gamma '} d_{\gamma}d_{\delta '} \bigg)
\label{Fourth kind}
\end{multline}
These terms describe two-particle cotunneling processes, e.g., the term $\propto d^\dagger_\delta d^\dagger_\gamma d_{\gamma'} d_{\delta'}$ describes tunneling from dots $\gamma'$, $\delta'$ to $\gamma$, $\delta$. Importantly, only the terms describing cotunneling between the two entangler dots 1, 2 and two detector dots $\alpha$, $\beta$ are relevant for the dot system dynamics, all other cotunneling processes have competing single-particle tunneling processes which dominate, similar to (ii) and (iii).

The total two-particle tunneling amplitude $t_{\beta \alpha 21}$ between the entangler and detector dots is given by summing up all terms in $S$ and $V$, yielding the relation $t_{\beta \alpha 21} = \frac{t_{\beta 1}t_{\alpha 2}}{\Delta E_{\beta \alpha}} - \frac{t_{\alpha 1}t_{\beta 2}}{\Delta E_{\alpha \beta}}$ in Eq.~(3) in the main text, where
\begin{widetext}
\begin{equation}
1/\Delta E_{\beta \alpha} = \frac{1}{2} \Big( \frac{1}{\epsilon_{1}+U_{12}-\epsilon_{\beta}-U_{2 \beta}} -\frac{1}{\epsilon_{2}+U_{2 \beta} - \epsilon_{\alpha}-U_{\alpha \beta}} + \frac{1}{\epsilon_{2}+U_{12} - \epsilon_{\alpha}-U_{1 \alpha}}  - \frac{1}{\epsilon_{1}+ U_{1 \alpha} - \epsilon_{\beta}- U_{\alpha \beta}}\Big)
\label{Two-particle cotunneling amplitudes}
\end{equation}
\begin{equation}
1/\Delta E_{\alpha \beta} = \frac{1}{2}\Big( \frac{1}{\epsilon_{2}+U_{12}-\epsilon_{\beta}-U_{1 \beta}} - \frac{1}{\epsilon_{1}+U_{1 \beta}-\epsilon_{\alpha}-U_{\alpha \beta}}+ \frac{1}{ \epsilon_{1}+U_{12} - \epsilon_{\alpha}-U_{2 \alpha}} - \frac{1}{\epsilon_{2}+U_{2 \alpha}-\epsilon_{\beta}-U_{\alpha \beta}}  \Big)
\label{Two-particle cotunneling amplitudes2}
\end{equation}
\end{widetext}

Denoting with $E_i$, $E_f$ and $E_m$ the initial, final and intermediate energies, respectively, of the system during the two-particle cotunneling, $1/\Delta E_{\beta \alpha}$, $1/\Delta E_{\alpha \beta}$ can each be written as $ \frac{1}{2}\sum_{m}(\frac{1}{E_i-E_m}-\frac{1}{E_m-E_f})$, where $m$ runs over the two possible intermediate states. We note that on two-particle resonance,  $E_f \equiv \epsilon_\alpha + \epsilon_\beta + U_{\alpha \beta} = \epsilon_1+\epsilon_2+U_{12} \equiv E_i$, this simplifies to $\sum_{m} \frac{1}{E_i-E_m}$, in accordance with standard second-order perturbation theory \cite{Averin&Nazarov}.

Using the two-particle tunneling amplitude in Eqs.~\eqref{Two-particle cotunneling amplitudes} and \eqref{Two-particle cotunneling amplitudes2}, the effective Hamiltonian can be written as
\begin{equation}
H_{\textrm{eff}} = H_0
+ \sum\limits_{\alpha \beta} \left( t_{\beta \alpha 2 1}d^\dagger_{\beta} d^\dagger_{\alpha} d_{2} d_{1}+t^*_{\beta \alpha 2 1}d^\dagger_{1} d^\dagger_{2} d_{\alpha} d_{\beta} \right)
\label{Final effective Hamiltonian}
\end{equation}
which is our Eq.~(2) in the main text.

We finally note that a full treatment, including the terms describing the leads and the dot-lead couplings in $H_0$ (not presented here), gives an additional level-broadening term in the denominator of $A_{\gamma \gamma'}$, proportional to the coupling strength $\Gamma_\alpha$, $\Gamma_\beta$ between a lead and its dot. Since these coupling strengths are much smaller than the differences between the single-particle energy levels of the dots, this level-broadening effect is negligible. Furthermore, cotunneling processes involving leads are suppressed in the regime considered, where the dot-lead couplings are weak compared to the energy differences between single-particle energy levels of the dots. Hence, to leading order, the effect of the leads can be neglected in the Hamiltonian.

\bibliographystyle{apsrev4-1}
\bibliography{sourcesSI}